\documentclass[a4paper,11pt]{article}
\usepackage{graphicx,amssymb,amstext,amsmath,mathabx,mathtools,esvect}
\usepackage{color}
\usepackage{floatflt}

\definecolor{cbl}{rgb}{0,0,1}                

\newcommand{\bc}{\begin{center}}
\newcommand{\ec}{\end{center}}
\def\ba#1{\begin{array}{#1}\displaystyle}
\newcommand{\ea}{\end{array}}

\newcommand{\beq}{\begin{equation}}
\newcommand{\eeq}{\end{equation}}
\newcommand{\beqa}{\begin{eqnarray}}
\newcommand{\eeqa}{\end{eqnarray}}

\newcommand{\bi}{\begin{itemize}}
\newcommand{\ei}{\end{itemize}}

\newcommand{\bra}{\langle}
\newcommand{\ket}{\rangle}

\newcommand{\Tr}{{\rm Tr}}

\newcommand{\TT}{{\cal T}}

\newcommand{\TTb}{\mathrm{T}\overline{\mathrm{T}}}
\newcommand{\bal}{\boldsymbol{\alpha}}
\newcommand{\bol}{\boldsymbol{0}}
\newcommand{\bel}{\boldsymbol{\beta}}

\usepackage{cite}

\definecolor{purple_nice}{rgb}{0.4,0.2,0.7}
\definecolor{fuel_blue}{RGB}{42,162,185}
\definecolor{YInMn_blue}{RGB}{46, 80, 144}
\definecolor{ultramarine}{RGB}{63, 0, 255}
\definecolor{KLEIN_blue}{rgb}{0, 0.18, 0.65}
\usepackage[linktocpage=true]{hyperref}
\hypersetup{
    colorlinks=true,
    linkcolor=YInMn_blue,
    citecolor=ultramarine,
    filecolor=fuel_blue,
    urlcolor=KLEIN_blue,
}

\makeatletter
\renewenvironment{abstract}{%
      \begin{center}%
        {\bfseries \normalsize\abstractname\vspace{\z@}}
      \end{center}%
      \quotation}
    {\endquotation}
\makeatother
\usepackage[normalem]{ulem}  

\begin{document}

\begin{titlepage}
\title{{Entanglement Entropy from Form Factors in $\TTb$-Deformed\\ Integrable Quantum Field Theories}}
\author{Olalla A. Castro-Alvaredo${}^{\heartsuit}$, Stefano Negro${}^\clubsuit$ and Fabio Sailis${}^{\diamondsuit}$\\[0.3cm]}
\date{{\small ${}^{\heartsuit,\diamondsuit}$ Department of Mathematics, City, University of London, 10 Northampton Square EC1V 0HB, UK\\[0.3cm]
${}^{\clubsuit}$ Center for Cosmology and Particle Physics, New York University, New York, NY 10003, U.S.A.\\[0.3cm]}}
\maketitle
\begin{abstract}
In two recent papers \cite{PRL,longpaper} we have proposed a program of study which allows us to compute the correlation functions of local and semi-local fields in generalised $\TTb$-deformed integrable quantum field theories. This new program, based on the construction of form factors, opens many avenues for future study, one of which we address in this paper: computing entanglement measures employing branch point twist fields. Indeed, over the past 15 years, this has become one the leading methods for the computation of entanglement measures, both in conformal field theory and integrable quantum field theory. Thus the generalisation of this program to $\TTb$-perturbed theories offers a promising new tool for the study of entanglement measures in the presence of irrelevant perturbations. In this paper, we show that the natural two-particle form factor solution for branch point twist fields in replica theories with diagonal scattering admits a simple generalisation to a solution for $\TTb$-perturbed theories. Starting with this solution, some of the known properties of entanglement measures in massive integrable quantum field theories can be generalised to the perturbed models. We show this by focusing on the Ising field theory. During the completion of this paper, we became aware of the recent publication \cite{chin} where the same problem has been addressed. 

\end{abstract}
\vfill

\noindent 
${}^{\heartsuit}$ o.castro-alvaredo@city.ac.uk\\
${}^{\clubsuit}$ stefano.negro@nyu.edu\\
${}^{\diamondsuit}$ fabio.sailis@city.ac.uk\\

\hfill \today
\end{titlepage}
\section{Introduction}
Quantum field theories perturbed by irrelevant perturbations, typically by the composite field $\TTb$ and higher spin versions thereof, have received a lot of attention in recent years. The earliest study of the composite field $\TTb$, particularly of its vacuum expectation value, was carried out in \cite{Zamolodchikov:2004ce} for generic 2D quantum field theory. A study of the matrix elements (form factors) of the same field in IQFT followed shortly afterwards \cite{Delfino:2004vc, Delfino:2006te}. Here ``composite" refers to its interpretation in conformal field theory as the leading field arising from the operator product expansion of the holomorphic ($T$) and anti-holomorphic ($\bar{T}$) components of the stress-energy tensor.

One of the most important results, especially for IQFTs, was established in \cite{Smirnov:2016lqw, Cavaglia:2016oda} where the $\TTb$ deformation was introduced for the first time and shown to be \emph{solvable}. This means that physical observables of interest, such as the finite-volume spectrum, the $S$-matrix \cite{Smirnov:2016lqw, Cavaglia:2016oda, Dubovsky:2017cnj} and the partition functions \cite{Dubovsky:2018bmo, Cardy:2018sdv, Datta:2018thy}, can all be determined exactly in terms of the corresponding undeformed quantities. The property of being solvable is also present in the generalised $\mathrm{T}\overline{\mathrm{T}}$ deformations, obtained by perturbing an IQFT\footnote{We stress that the $\mathrm{T}\overline{\mathrm{T}}$ deformation of \emph{any} 2D field theory -- whether integrable or not -- is solvable. On the other hand, the generalised $\mathrm{T}\overline{\mathrm{T}}$ deformations mentioned in the text require the existence of higher spin conserved currents, so they can usually only be performed in IQFTs.} by composite operators constructed from higher-spin conserved currents \cite{Conti:2019dxg, Hernandez-Chifflet:2019sua}. Indeed, a (generalised) $\mathrm{T}\overline{\mathrm{T}}$-deformation in an IQFT has the effect of modifying the two-body scattering matrix by a particular type of CDD factor \cite{Smirnov:2016lqw, Cavaglia:2016oda, Camilo:2021gro, Cordova:2021fnr}, as shown in equation (\ref{2}). This feature is extremely important because in IQFT the two-body scattering matrix fully characterises all scattering processes of the theory and can be almost entirely determined by a consistency procedure known as \emph{bootstrap} \cite{Zamolodchikov:1978xm, Zamolodchikov:1989hfa}. The ``almost'' here refers to the so-called \emph{CDD ambiguity} \cite{Zamolodchikov:1978xm}: the boostrap equations admit a ``minimal solution'' (i.e., a solution having the minimal set of singularities) which can then be dressed by an arbitrary number of CDD factors \cite{Castillejo:1955ed}. These are trivial solutions of the $S$-matrix bootstrap equations. They can modify the $S$-matrix in non-trivial ways, but they do not include poles in the physical strip, thus leaving the spectrum of the theory intact. Recent work by the present authors \cite{PRL,longpaper} has shown that solvability, as defined above, is also an attribute of the matrix elements of local and semi-local fields. That is, these can be determined exactly by following the standard form factor program \cite{Karowski:1978vz,smirnov1992form} and, like the $S$-matrix, are related to their underformed counterpart by a multiplicative factor.

\medskip
Let us recall the expression of the (deformed) scattering matrix. In general, for a theory with diagonal scattering and a single particle spectrum, perturbation by $\TTb$ and its higher spin descendants leads to a new scattering matrix of the form \beq
S_{\boldsymbol{\alpha}}(\theta) := S_{\boldsymbol{0}}(\theta) \Phi_{\boldsymbol{\alpha}}(\theta)\, \quad {\mathrm{with}} \quad \log \Phi_{\boldsymbol{\alpha}}(\theta)=-i\sum\limits_{s\in \mathcal{S}}\alpha_s m^{s+1} \sinh(s\theta)\,.
\label{2}
\eeq 
Here $S_{\boldsymbol{0}}(\theta) $ is the underformed exact scattering matrix and $\bal=\{\alpha_1,\alpha_2,\ldots\}$ is a set of coupling constants labelled by the spin $s\in \mathcal{S}$. The set $\mathcal{S}$ represents the values of the spin of local conserved quantities which are typically odd integers. The set of such integers depends on the specific theory under consideration \cite{Zamolodchikov:1989hfa,Negro:2016yuu}.  Note that the sum can in principle contain infinitely many terms but in such a case one should be careful to determine the convergence properties of the sum in the right-most side of \eqref{2}. In general, for $\bal$ of infinite cardinality, the quantity $\log\Phi_{\bal}(\theta)$ will split into a convergent sum and an additional term that introduces new poles outside the physical strip (see \cite{Camilo:2021gro} for more details). Consequently, the properties of the associated $S$-matrix and form factors can be radically different. In this work we will focus on a finite number of terms, often a single term with $s=1$, corresponding to the $\TTb$ perturbation. 

It is easy to see that the CDD factors $\Phi_{\boldsymbol{\alpha}}(\theta)$ in \eqref{2} automatically satisfy all $S$-matrix consistency equations. However it presents a very uncommon double-exponential dependence on the rapidity that changes the $S$-matrix asymptotic behaviour and has a stark effect on the theory's RG flow. Generally speaking, whereas  the IR regime is left unaltered by the perturbation, in the short-distance limit the theory displays unusual features, which are incompatible with the existence of a UV CFT \cite{Dubovsky:2012wk, Dubovsky:2013ira, Dubovsky:2017cnj}. Some of these features are also reflected on the behaviour of correlation functions, as studied in \cite{PRL,longpaper}. The first work on form factors and correlation functions in irrelevantly deformed IQFT \cite{sgMuss} actually pre-dates the introduction of the $\TTb$ deformation and focus on a very particular deformation of the sinh-Gordon model, where $\log\Phi_{\bal}(\theta) = i\pi$.

Beyond the works mentioned so far,  generalised $\mathrm{T}\overline{\mathrm{T}}$ deformations and their properties have been studied from a multitude of viewpoints, hence the vast literature already in existence. The problem has been studied in the context of 2D classical and quantum field theory \cite{Conti:2018jho, Conti:2018tca, Dubovsky:2023lza}, in 1D quantum-mechanical systems 
\cite{Gross:2019ach, Gross:2019uxi, ste_zeta}. Some generalizations to higher dimensions have been proposed a few years ago \cite{Cardy:2018sdv, Bonelli:2018kik, Taylor:2018xcy} and, more recently, a promising approach that relies on the geometric picture of \cite{Conti:2018tca,Dubovsky:2018bmo} has been advanced in \cite{Conti:2022egv}. Further studies \cite{Aramini:2022wbn} have proven the compatibility of the ODE/IM correspondence \cite{Dorey:2019ngq} and the $\mathrm{T}\overline{\mathrm{T}}$ deformation. A particularly fertile approach for IQFTs has been the thermodynamic Bethe ansatz (TBA). Indeed, the first TBA analysis on models with S-matrices of the type \eqref{2} (for the $\TTb$ case) were performed in \cite{Dubovsky:2012wk, Caselle:2013dra}, before the ``official'' introduction of the $\mathrm{T}\overline{\mathrm{T}}$ deformation in \cite{Smirnov:2016lqw, Cavaglia:2016oda}. Further studies followed in \cite{Cavaglia:2016oda, Conti:2019dxg, Hernandez-Chifflet:2019sua, Camilo:2021gro, Cordova:2021fnr, LeClair:2021opx, LeClair:2021wfd, Ahn:2022pia}. Other approaches used include perturbed conformal field theory \cite{Guica:2017lia, Cardy:2018sdv, Cardy:2019qao, Aharony:2018vux, Aharony:2018bad, Guica:2020uhm, Guica:2021pzy, Guica:2022gts}, string theory \cite{Baggio:2018gct, Dei:2018jyj, Chakraborty:2019mdf, Callebaut:2019omt}, holography \cite{McGough:2016lol, Giveon:2017nie, Gorbenko:2018oov, Kraus:2018xrn, Hartman:2018tkw, Guica:2019nzm, Jiang:2019tcq,Jafari:2019qns}, quantum gravity \cite{Dubovsky:2017cnj, Dubovsky:2018bmo, Tolley:2019nmm, Iliesiu:2020zld, Okumura:2020dzb, Ebert:2022ehb} and out-of-equilibrium conformal field theory \cite{Medenjak:2020ppv, Medenjak:2020bpe}. A natural counterpart of the $\TTb$ perturbation for spin chains was studied in \cite{PJG, Marchetto:2019yyt} and shown to correspond to a specific type of integrable long-range spin chain studied a decade before \cite{Bargheer:2008jt,Bargheer:2009xy}. Finally the generalised hydrodynamics (GHD) approach \cite{Cardy:2020olv} has provided a new interpretation of the effect of irrelevant perturbations on 1D quantum theories, namely as endowing the particles of the original models with a finite, positive or negative length. 

{
Of particular relevance to the present paper are works which have focused on entanglement measures \cite{Donnelly:2018bef, Chen:2018eqk, Jeong:2019ylz, He:2019vzf, Allameh:2021moy, Jiang:2023ffu}. A perturbative approach for free theories was used in \cite{Ashkenazi:2023fcn} 
and led to some interesting conclusions regarding the leading contribution to the entanglement entropies in free boson and free fermion theories. More recently the work
\cite{chin} has addressed exactly the same problems we consider here. In fact,the present work was very nearly completed when \cite{chin} appeared. Despite the large overlap between our work and \cite{chin}, we have found important differences in some of the main results. We will discuss and highlight those differences in detail in the core of the paper.} 

This paper extends the work in \cite{longpaper,PRL} to deal with a particular type of symmetry field, the branch point twist field (BPTF). The name BPTF, their definition as symmetry fields associated with cyclic permutation symmetry in replica theories, and their application to entanglement measures in massive IQFTs all hail back to the work \cite{entropy}. The relationship between entanglement measures and correlators of fields associated with conical singularities was first discussed in \cite{Calabrese:2004eu}, although the fields introduced there were distinct from the BPTF. The extension considered here is non-trivial, because the results of \cite{PRL} were only applicable to local and semi-local fields, where semi-local meant specifically fields with factor of local commutativity $\gamma=\pm 1$ \cite{Yurov:1990kv}. Indeed, these are the types of fields we find in the Ising field theory, which we studied in great detail in \cite{longpaper}. This means that, in order to study the form factors of BPTFs we need to consider the form factor program anew and find solutions to the form factor equations which might a priori  have a different structure from those found in \cite{PRL,longpaper}. For now, we will just focus on the two particle form factor (the zero and one-particle form factors are constant for spinless fields like the BPTF), leaving higher particle solutions for future study. As we will see, the two-particle contribution can already give us valuable insights into some properties of entanglement.

\medskip 
The paper is organised as follows: In Section 2 we review the form factor program for BPTFs and propose a solution for the two-particle form factor, based on a new minimal form factor. We then study the $\Delta$-sum rule in the two-particle approximation. 
In Section 3 we introduce some basic definitions and results regarding the entanglement entropy and show that the next-to-leading order correction to the von Neumann entropy of a large subsystem is identical to that found for undeformed theories. Assuming that the $\Delta$-sum rule provides a good description of how the vacuum expectation value of the BPTF scales as a function of mass, we show that the leading contribution to entanglement (i.e. its saturation value in a gapped theory) is modified in the perturbed theory. 
We conclude in Section 4. 

\section{Form Factor Program for Branch Point Twist Fields}
It has been known for a long time that fields associated to internal symmetries in quantum field theory can be characterised by their equal-time exchange relations with respect to other local fields and that these in turn lead to twist fields sitting at the origin of branch cuts in space-time (see e.g. \cite{IZ,Schroer,BABE} for the elaboration of these ideas in the context of the Ising field theory). The fields we are interested in here, the BPTFs, arise naturally in replica theories which consist of $n$ identical, non-interacting copies of some chosen model. In such cases, there is a large amount of symmetry in the resulting model (in fact, under all elements of the permutation group), including cyclic permutation symmetry, which is the symmetry that turns out to be relevant in the context of entanglement. As a result of the replica process, all the fields that were present in the original (non-replicated) model acquire copy labels. In other words, if the chosen model had a local field $\phi(\boldsymbol{x})$, the replica theory will posses a family $\phi_j(\boldsymbol{x})$ with $j=1,\ldots, n$ and the identification $j\equiv j+n$. In this context 
the BPTF $\TT$ and its hermitian conjugate $\TT^\dagger$ satisfy the equal-time exchange relations
\beq
\phi_j (x)\TT(y)=\left\{ \begin{array}{cc}
\TT(y) \phi_{j+1}(x)& y>x\\
\TT (y) \phi_j(x) & y<x
\end{array}\right. \quad \mathrm{and}\quad \phi_j (x)\TT^\dagger(y)=\left\{ \begin{array}{cc}
\TT^\dagger(y) \phi_{j-1}(x)& y>x\\
\TT^\dagger(y) \phi_j(x) & y<x
\end{array}\right.\;.
\label{T}
\eeq 
From these exchange relations, equations for matrix elements of $\TT$ and $\TT^\dagger$ can be derived. Let us define
\beq 
F_k^{\mu_1\ldots\mu_k}(\theta_1,\ldots,\theta_k;n):=\bra 0| \TT(0)|\theta_1,\ldots,\theta_k\ket_{\mu_1\ldots\mu_k}\,,\label{FF}
\eeq 
to be the $k$-particle form factor of $\TT$, with  $|\theta_1,\ldots,\theta_k\ket_{\mu_1,\ldots,\mu_k}$ an in-state characterised by rapidities $\theta_i$ and quantum numbers $\mu_j$. Typically, each quantum number carries two sets of information: the particle quantum number and its copy number; in short $\mu_j:=(a,j)$. However if the theory has a single particle type, as we consider here, we can think of $\mu_j:=j$ as being simply the copy number. From the relations \eqref{T}, the form factor equations for diagonal scattering matrices can be derived and were first written in \cite{entropy}:
\beqa 
    F_k^{\ldots \mu_i \mu_{i+1}\ldots }(\ldots, \theta_i, \theta_{i+1},\ldots; n)&=& S_{\mu_i \mu_{i+1}} (\theta_{i\,i+1}) F_k^{ \ldots  \mu_{i+1} \mu_i\ldots}  (\ldots,  \theta_{i+1},\theta_i\ldots ; n)\,, \label{ex}\\     
    F_k^{\mu_1 \mu_2 \ldots \mu_k }( \theta_1 + 2 \pi i,\ldots, \theta_k;n)&=& F_k^{\mu_2 \ldots \mu_n \hat{\mu}_1} (\theta_2, \ldots, \theta_k, \theta_1;n)\,,  \label{cross} \\
    \frac{1}{i} \underset{\Bar{\theta}_0=\theta_0}{\text{Res}} F_{k+2}^{ \Bar{\mu} \mu \mu_1  \ldots \mu_k} (\Bar{\theta}_0+ i \pi, \theta_0, \theta_1 , \ldots, \theta_k;n) &=& F_k^{\mu_1 \ldots \mu_k}(\theta_1 ,  \ldots, \theta_k;n)\,, \label{kre} \\
   \frac{1}{i} \underset{\Bar{\theta}_0=\theta_0}{\text{Res}} F_{k+2}^{\Bar{\mu} \hat{\mu} \mu_1  \ldots \mu_k} (\Bar{\theta}_0+ i \pi, \theta_0, \theta_1 , \ldots, \theta_k; n)&=& - \prod_{i=1}^k S_{\mu \mu_i}(\theta_{0i}) F_k^{\mu_1 \ldots \mu_k}(\theta_1 ,  \ldots, \theta_k;n) \label{kre2}\,,
\eeqa 
where $\hat{\mu}_i:=(a,i+1)$ indicates a particle label with the same quantum numbers as $\mu$ but copy number shifted by one  and $\bar{\mu}_i:=(\bar{a},i)$ indicates a particle that is conjugate to the particle of index $\mu$ in the same copy. Although the kinematic residue equation for branch point twist fields now splits into two separate equations (\ref{kre})-(\ref{kre2}), the second can be derived from the first by employing Watson's equations (\ref{ex})-(\ref{cross}). This means that typically only the equations (\ref{ex})-(\ref{kre}) are employed in the solution procedure. Furthermore, for simplicity, these can be specialised to particles in the same copy since once those form factors are known, all others can be obtained by employing (\ref{ex})-(\ref{cross}). The form factors of $\TT^\dagger$ are also related to those of $\TT$ in a simple way, as discussed in \cite{entropy} and \cite{review}.

As usual, the solution to these equations starts with solving equations for two-particle form factors. We now discuss how this is done and how the known solutions generalise in the presence of $\TTb$ perturbations. In this case, all equations above still hold, but, in a theory with a single particle type, the scattering matrices $S_{\mu_i \mu_j}(\theta)$ are replaced by $(S_{\bal}(\theta))^{\delta_{ij}}$ and therefore the form factors must now be functions of $\bal$.

\subsection{The Minimal Form Factor}
Since we want to deal with deformed theories, let us slightly modify our notation (\ref{FF}) and introduce explicitly the dependence on the couplings $\bal$. In addition, since BPTFs are spinless we know their two-particle form factor will only depend on rapidity differences, hence a single variable $\theta$. We will then call 
\beq 
F_2^{jp}(\theta;n,\bal)\;,
\eeq 
the two-particle form factor of $\TT$ in the deformed theory for two particles of the same type, labelled by distinct copy indices $j,p$. The minimal part of this form factor will be denoted by 
\beq 
F_{\rm min}^{jp}(\theta;n,\bal)\;,
\eeq 
and will be a solution of the two-particle form factor axioms without poles in the physical sheet. It satisfies
\begin{equation}
        F_{\text{min}}^{pj}(\theta;n,\bal)= F_{\text{min}}^{jp}(-\theta;n, \bal) (S_{\bal}(\theta))^{\delta_{jp}}=F_{\text{min}}^{j\, p +1}(2 \pi i-\theta;n,\bal)\;, \qquad \forall j,p\;, \label{2par}
    \end{equation}
where the delta function implements the correct exchange relation (\ref{ex}) in which particles in distinct copies do not interact whereas particles in the same copy interact in a model specific manner.  Using these relations as well as the fact that all copies are identical, we can show 
\begin{flalign}
    & F_{\text{min}}^{\mathcal{T}| i\,i+p }(\theta;n,\bal)= F_{\text{min}}^{\mathcal{T}| j\, j +p} (\theta,n)\;, \qquad \forall \; i,j,p\;,  \\
    & F_{\text{min}}^{\mathcal{T}| 1j}(\theta;n,\bal)= F_{\text{min}}^{\mathcal{T}| 11}\left( 2 \pi(j-1)i - \theta;n,\bal\right)\;, \qquad \forall \; j\neq 1\;.
\end{flalign}
and the last equation allows us to write a similar set of equations to (\ref{2par}) for form factors involving particles in the same copy only. This gives the relations
\begin{flalign}
    f_n^{\bal}(\theta) = f_n^{\bal}(-\theta) S_{\bal}(\theta)= f_n^{\bal}(- \theta + 2 \pi i n)\;.  \label{minFFeqs}
\end{flalign}
which imply that the minimal form factor
\beq 
F_{\text{min}}^{11}(\theta;n,\bal) :=f_n^{\bal}(\theta)\;,
\eeq 
must not have any pole in the extended strip $\mathrm{Im}(\theta) \in [0,2 \pi n]$. For $\bal=\bol$ the relations (\ref{minFFeqs}) can be typically solved in a very similar way as for local fields and the method has been explained in many places including \cite{entropy,review}. The resulting function $f_n^{\bol}(\theta)$ can be usually expressed in terms of an integral representation which is determined by the scattering matrix. For free theories however, this function is very simple. In particular, for the Ising field theory it is just
\beq
f_n^{\bol}(\theta)=-i\sinh\frac{\theta}{2n}\;.
\eeq 
Since the equations (\ref{minFFeqs}) are linear, this means that their solution will typically factorise into $f_n^{\bol}(\theta)$ (the undeformed solution) and another function $\varphi_n^{\bal}(\theta)$ which satisfies
\beq 
    \varphi_n^{\bal}(\theta) = \varphi_n^{\bal}(-\theta) \Phi_{\bal}(\theta)= \varphi_n^{\bal}(- \theta + 2 \pi i n)\;.  \label{mindef}
\eeq 
A simple family of solutions to this equation can be found similarly as for local fields \cite{PRL,longpaper}. It is a simple generalisation to replica theories of the solutions found there, and it takes the form 
\beq
        \varphi_n^{\bal}(\theta)=\exp\left({\frac{\theta-i\pi n}{2 \pi n } \sum_{s\in\mathcal{S}} \alpha_s m^{s+1} \sinh(s\theta) }\right)\;.
\label{cdd}
\eeq 
As for local and semi-local fields, the minimal form factor itself contains possible ``CDD factors'', that is factors that are both even and $2\pi i n$ periodic in the variable $\theta$. The most general combination of such factors takes the form 
\beq 
C^{\bel}(\theta)=\exp\left(\sum_{s\in\mathcal{S}'} \beta_s m^{s+1} \cosh\frac{s\theta}{n}\right)\;.
\label{cdd2}
\eeq 
 where $\mathcal{S}'$ may be a different set of integers than $\mathcal{S}$. {Our solutions (\ref{cdd}) and (\ref{cdd2}) are identical to those found in \cite{chin}. It is however useful to discuss a little bit why we choose (\ref{cdd}) to involve only hyperbolic functions of $s\theta$ whereas (\ref{cdd2}) involves hyperbolic functions of $\frac{s \theta}{n}$. Indeed, if $s$ are odd integers, which is usually the case, we could replace $\sinh(s\theta)$ by $\sinh\frac{s\theta}{n}$ and still have a solution to all required equations. The reason for choosing (\ref{cdd}) as done here is two-fold: it is natural to expect that the minimal form factor satisfies} {
 \beq 
 f_n^{\bal}(\theta) (f_1^{\bal}(\theta))^* \in \, \mathbb{R} \quad \rm{for} \quad \theta \in \mathbb{R}\,\quad {\rm{and}}\quad {\rm n}\neq 1\;,
 \label{real}
 \eeq 
 as otherwise any correlators of $\TT$ with a local or semi-local field in the theory would give complex results. The property (\ref{real}) holds for $\bal=\bol$. This property is for instance essential in the following section where we study the $\Delta$ sum rule. Without (\ref{real}), the output of the $\Delta$-sum rule presented below would not be a real number, even in the two-particle approximation. The choice of  $\varphi_n^{\bal}(\theta)$ is also supported by fundamental properties of the $S$-matrices and minimal form factors of IQFTs which we plan to discuss in more detail in a further publication.} 
 
Furthermore, the solution (\ref{cdd2}) is the most general CDD factor that we can write and of course it includes all terms of the form $\cosh(k\theta)$ when $s=k n$. Terms of the form $\cosh(k\theta)$ seem indeed more natural as they are the one-particle eigenvalues of higher local conserved quantities. It is however also possible to interpret the functions $\cosh\frac{s\theta}{n}$ as the one-particle eigenvalues of quasi-local charges of fractional spin, which play an important role in the description of generalised Gibbs ensembles (see e.g.\cite{gge,Fioretto,gge2}).
 
 In summary, we have then that any diagonal theory with a single-particle spectrum has minimal form factor
    \beq 
f_n^{\bal}(\theta) = f_n^{\bol}(\theta) \varphi_n^{\bal}(\theta)C^{\bel}(\theta)\;.
    \eeq 
In the Ising model this reads explicitly
 \beq 
f_n^{\bal}(\theta) = -i \sinh\frac{\theta}{2n} \exp\left({\frac{\theta-i\pi n}{2 \pi n } \sum_{s\in 2\mathbb{Z}+1} \alpha_s m^{s+1} \sinh(s\theta) }+ \sum_{s\in \mathcal{S}'} \beta_s m^{s+1} \cosh\frac{s\theta}{n}\right)\;.
    \eeq 
    If we now include the relevant kinematic poles and normalisation so as to ensure that the two-particle versions of equations (\ref{kre}) and (\ref{kre2}) are also satisfied, we obtain the solution
\begin{equation}
    F_2^{11}(\theta; n,\bal) = \frac{\langle \mathcal{T} \rangle_n^{\bal} \,\sin\frac{\pi}{n}}{2n \sinh{[\frac{i \pi +\theta}{2n}]} \sinh{[\frac{i \pi -\theta}{2n}}]} \frac{f_n^{\bal}(\theta) }{f_n^{\bal}(0)}\;.
    \label{22}
\end{equation}
where $\langle \mathcal{T} \rangle_n^{\bal} :=F_0(n,\bal)$. Note that the factor $\sin\frac{\pi}{n}$ ensures that $F_2^{11}(\theta; 1,\bal)=0$ which is a consistency condition for the BPTF (i.e. it becomes the identity field for $n=1$). Apart from the minimal form factor, this form factor has exactly the usual form \cite{entropy}. 

It is worth highlighting however, that this solution still needs to be tested against solutions to the higher particle form factor equations, which are harder to obtain.  We mean by this that the two-particle form factor needs to be consistently recovered from the kinematic residue of the four-particle form factor. Until this four-particle form factor is computed some uncertainty as to the exact form of $F_2^{11}(\theta;n,\bal)$ remains. This is similar to the behaviour we found in \cite{longpaper} for the form factors of the trace of the stress-energy tensor $\Theta$. In that case, we showed that consistency with higher-particle form factor equations requires the standard two-particle form factor solution to be multiplied by an additional factor $h(\theta;\bal)$ with properties
\beq 
h(\theta;\bol)=h(i\pi;\bal)=1\;.
\eeq 
Such a factor could be present in our case too. If so, it would need to be also an even, $2\pi i n$-periodic function of $\theta$.
A more in-depth study of the form factor equations is required to settle this point. This is work in progress. For now, we will work with the solution \eqref{22}, the same found in \cite{chin}. 

\subsection{The $\Delta$-Sum Rule}
One of the typical uses and consistency check for form factors is the computation of the conformal dimension of the operator in the underlying UV theory via the $\Delta $-sum rule \cite{DSC}. Given that the UV limit of $\TTb$-deformed theories is not a CFT in the standard sense, the applicability and meaning of this rule are still to be understood. However, we found in \cite{longpaper} that for $\alpha_{s}<0$ where $s$ is the highest spin involved in the sum (\ref{2}), both the form factor expansion of two-point functions and their integration according to the identity below are convergent and predict qualitatively similar properties. The result is a function of $\bal$ with properties that are compatible with the power-law scaling of two-point functions that we observed in \cite{PRL,longpaper}. Even if the form factors of the BPTF are different from those of local and semi-local fields previously found, power-law scaling will follow through due to the exponentially decaying factors $|\varphi_n^\alpha(\theta)|^2$ that enter the form factor expansion of the two-point function $\bra \TT(0)\TT(r)\ket_n^\alpha$. For the same reason, the $\Delta$-sum rule in the two-particle approximation is also given by a convergent integral. 

Here we will consider the Ising model, but in general the $\Delta$-sum rule can be expressed as \cite{RGflow}
\begin{flalign}
    \Delta_{n}^{\bal} &= -  \frac{1}{2 \langle \mathcal{T} \rangle_n^{\bal}} \int_0^{\infty} dr \, r \langle \Theta(r) {\mathcal{T}}^\dagger(0)
 \rangle_n^{\bal} =  \nonumber \\
 &= -  \frac{1}{2 \langle \mathcal{T} \rangle_n^{\bal}} \sum_{k=1}^{\infty} \sum_{\mu_1 \dots \mu_k} \int_{-\infty}^{\infty} \cdots \int_{-\infty}^{\infty}  \frac{d\theta_1 \dots d\theta_k}{k! (2 \pi)^k (\sum_{i=1}^k m_{\mu_i} \cosh{(\theta_i)})^2} \times \label{Deltasum} \\
 & \quad \, \times F_k^{\Theta | \mu_1 \dots \mu_k}(\theta_1 ,  \dots, \theta_k;\bal) ( F_k^{\mathcal{T} |\mu_1 \dots \mu_k}(\theta_1 ,  \dots, \theta_k; n,\bal) )^* \;. \nonumber 
 \end{flalign}
The series in (\ref{Deltasum}) is rapidly convergent for $\alpha_s<0$ and we expect the main contribution to come from the first few terms. We can then approximate (\ref{Deltasum}) with the two-particle contribution only, and after some simplifications we find
\begin{equation}
    \Delta_{n}^{\bal} \approx -  \frac{n}{32 \pi^2 m^2 \langle \mathcal{T} \rangle_n^{\bal}} \int_{-\infty}^{\infty} d \theta \frac{F_2^{\Theta | 11}(\theta;\bal) (F_2^{\mathcal{T} |11}(\theta; n,\bal))^*}{\cosh^2{(\theta/2)}}\;.
\end{equation}
If we now consider the $\TTb$ perturbation, corresponding to having only $\alpha_1:=\alpha\neq 0$ and all $\bel=\bol$ we then have the form factor formulae 
\begin{flalign}
    & F_2^{\Theta | 11}(\theta;\alpha)= - 2 \pi i m^2 \left|\frac{\sin(\frac{\alpha}{2}\sinh\theta)}{\frac{\alpha}{2}\sinh\theta}\right| \sinh\frac{\theta}{2} e^{-\frac{\alpha m^2}{2 \pi}(i \pi -\theta) \sinh \theta}\;, \\
    & F_2^{\mathcal{T} |11}(\theta; n,\alpha)= - \frac{i \langle \mathcal{T} \rangle_n^{\bal} \sinh{(\frac{\theta}{2n})} \cos{(\frac{\pi}{2n})} }{n \sinh{[\frac{ i \pi -\theta }{2n}]} \sinh{[\frac{i \pi+ \theta }{2n}]} } e^{-\frac{\alpha m^2}{2 \pi }(i \pi -\frac{\theta}{n}) \sinh\theta}\;, \label{twistFF}
\end{flalign}
where the first formula was obtained in \cite{longpaper,PRL}.
We find\footnote{In \cite{chin} the authors chose to ignore the factor $\left|\frac{\sin(\frac{\alpha}{2}\sinh\theta)}{\frac{\alpha}{2}\sinh\theta}\right|$ arguing that it does not influence the final result too much. This is indeed true, but it seems a peculiar reasoning nonetheless.}
\begin{equation}
     \Delta_{n}^{\bal} = -  \frac{1}{16 \pi} \int_{-\infty}^{\infty}  d \theta \frac{\cos{(\frac{\pi}{2n})} \sinh{(\frac{\theta}{2n})} \tanh\frac{\theta}{2}}{\sinh{[\frac{ i \pi -\theta }{2n}]} \sinh{[\frac{i \pi+ \theta }{2n}]} \cosh\frac{\theta}{2} } \left|\frac{\sin(\frac{\alpha}{2}\sinh\theta)}{\frac{\alpha}{2}\sinh\theta}\right| e^{\frac{\alpha m^2}{2 \pi}( \frac{n+1}{n}) \theta\sinh\theta}\;.
    \label{DIsing}
\end{equation}
For $\alpha=0$ the integral can be computed exactly as shown in \cite{entropy}, and it gives the exact CFT formula \cite{kniz,orbifold}
\beq 
  \Delta_{n}^{\bol}=\frac{1}{48}\left(n-\frac{1}{n}\right)\;.
\label{exact}
\eeq 
In \cite{entropy} it was shown that the integral (\ref{DIsing}) can be computed exactly for $\bal=0$. This is because in that case the integrand only picks up a sign under the shift $\theta \mapsto \theta + 2\pi i n$. Hence by Cauchy's theorem, the sum of the residues of the poles that cross the line under the shift give exactly twice the dimension $\Delta_n^{\bol}$. In the present case, unfortunately the presence of the exponential spoils this property. Note that this fact has been overlooked in \cite{chin} where the integral was computed as a sum of residues leading to an expression that is quadratic in $\alpha$. That expression captures some of the correct features of the integral (such as its linear dependence on $n$ for large $n$) however it is not correct. This can be seen in several ways however the simplest test is the observation that (\ref{DIsing}) is zero for $n=1$ (due to the factor $\cos\frac{\pi}{2n}$) and this is also the case for the integral formula employed in \cite{chin}. On the other hand, the ``exact" formulae for the dimension given in \cite{chin} are not zero at $n=1$.  

\begin{figure}[h!]
\begin{center}
	\includegraphics[width=7.5cm]{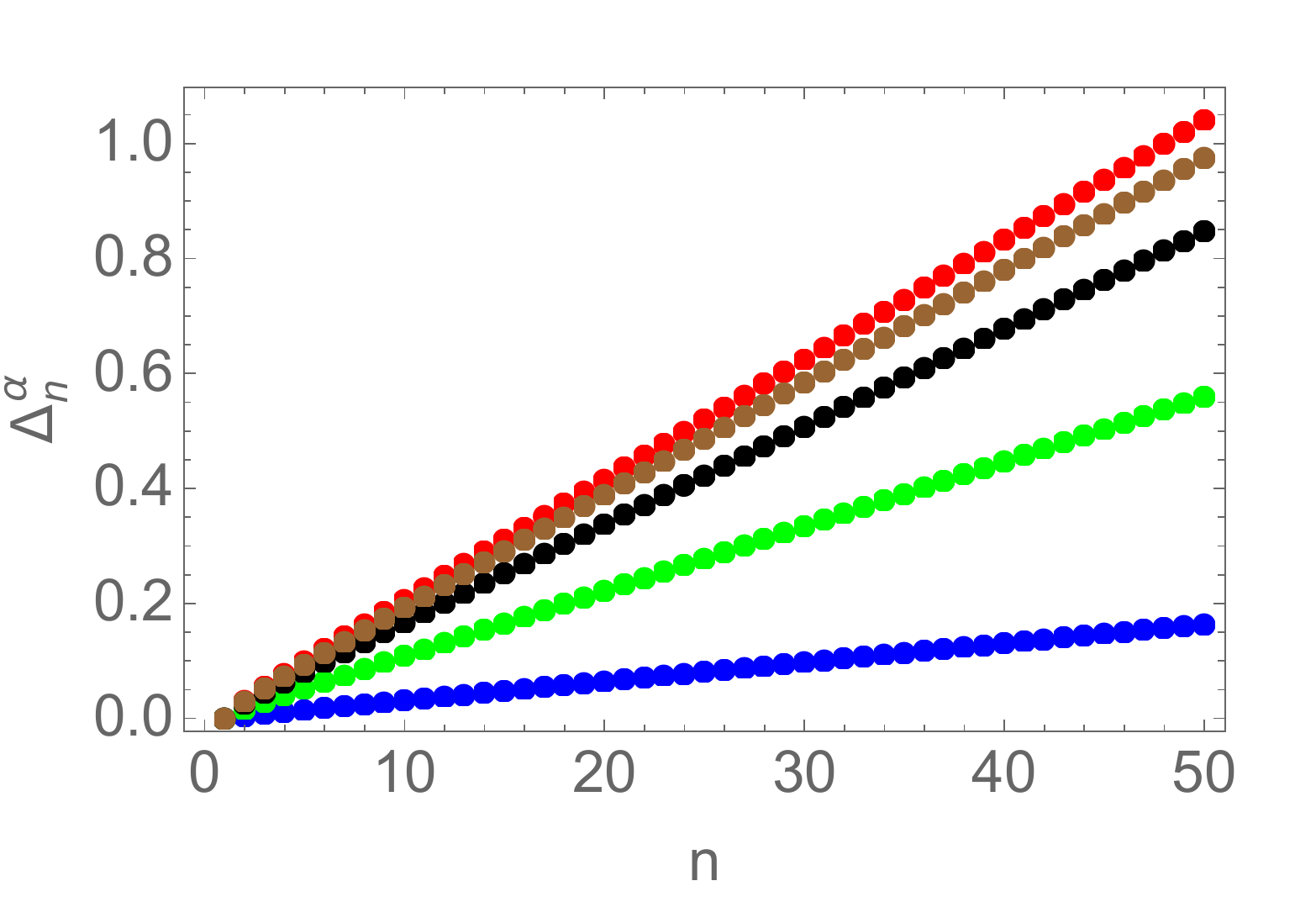}
 \includegraphics[width=7.5cm]{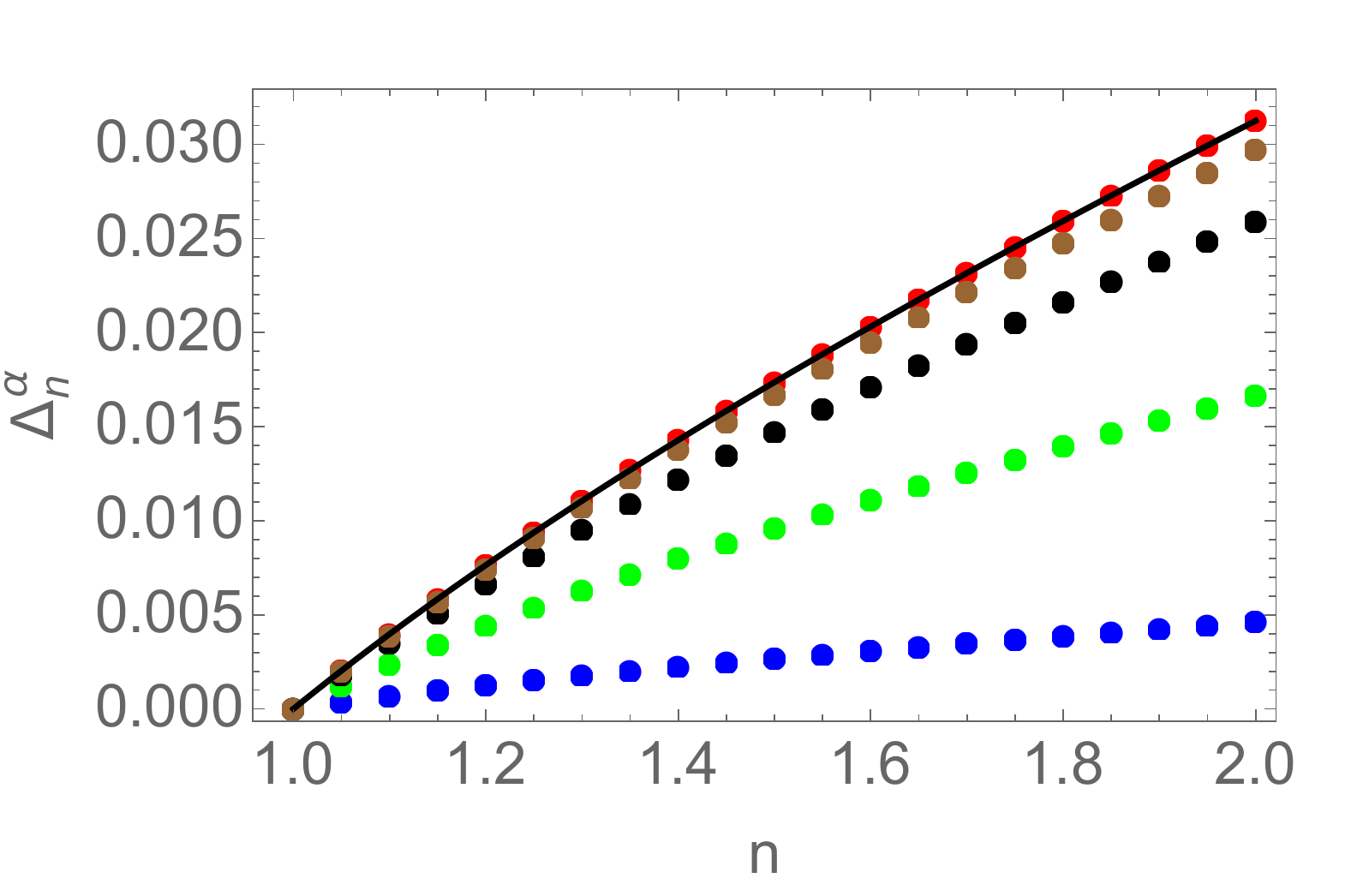}

				    \caption{The function (\ref{DIsing}) for $\alpha=0, -0.001, -0.01, -0.1$ and $-1$, with the value 0 corresponding to the red dots and the value $1$ corresponding to the blue dots. The exact formula (\ref{exact}) is also plotted as a solid black line, which fits the red dots perfectly. The figure on the right is a blow up of the small $n$ region of the left figure. All functions are linear for $n$ large and the slopes are: $b(0)=\frac{1}{48}=0.0208333$, $b(-0.001)=0.0195262$, $b(-0.01)=0.0169936$, $b(-0.1)=0.0112533$ and $b(-1)=0.00331503$.}
				     \label{figure1}
    \end{center}
    \end{figure}

\begin{figure}[h!]
\begin{center}
	\includegraphics[width=7.5cm]{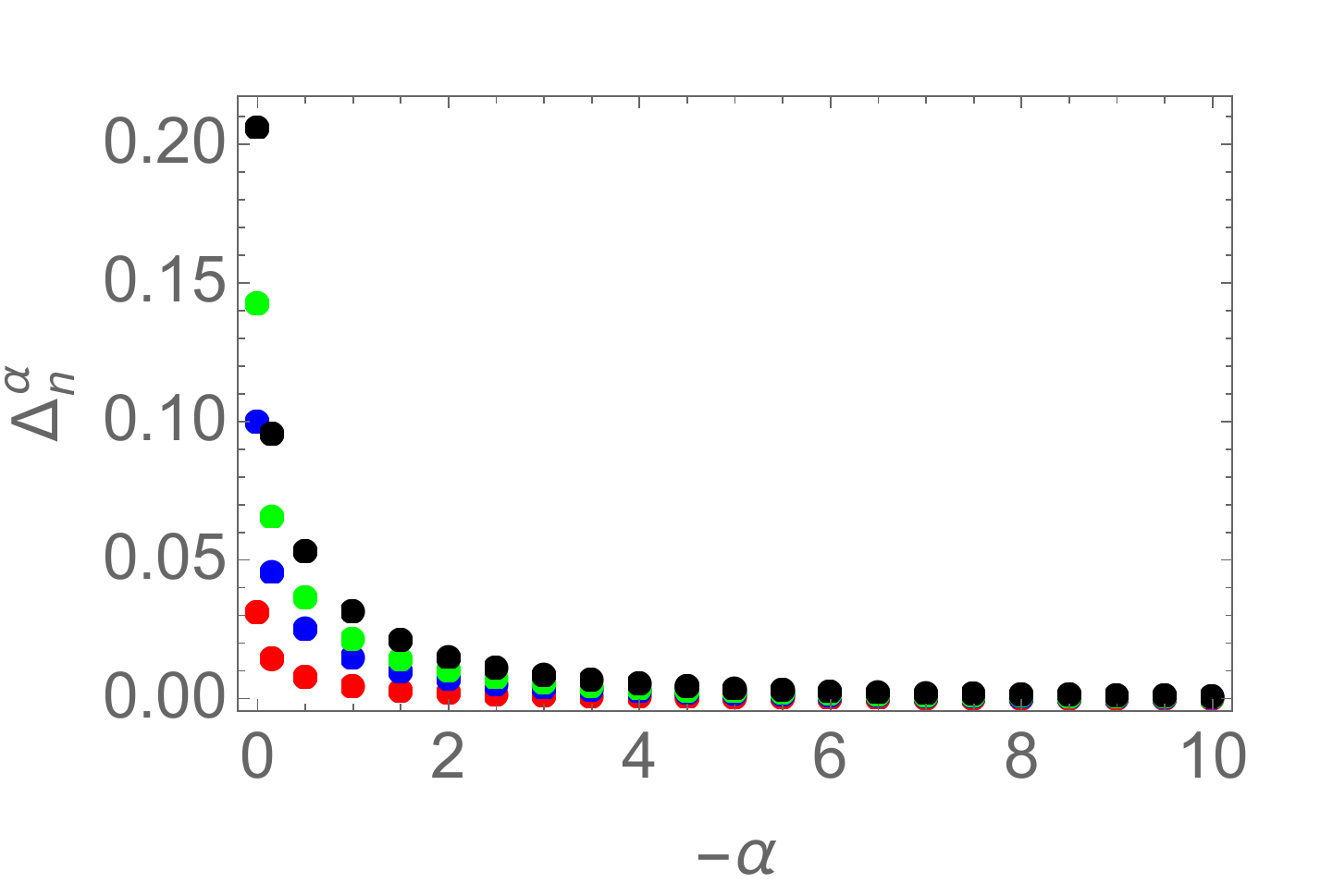}
 \includegraphics[width=7.5cm]{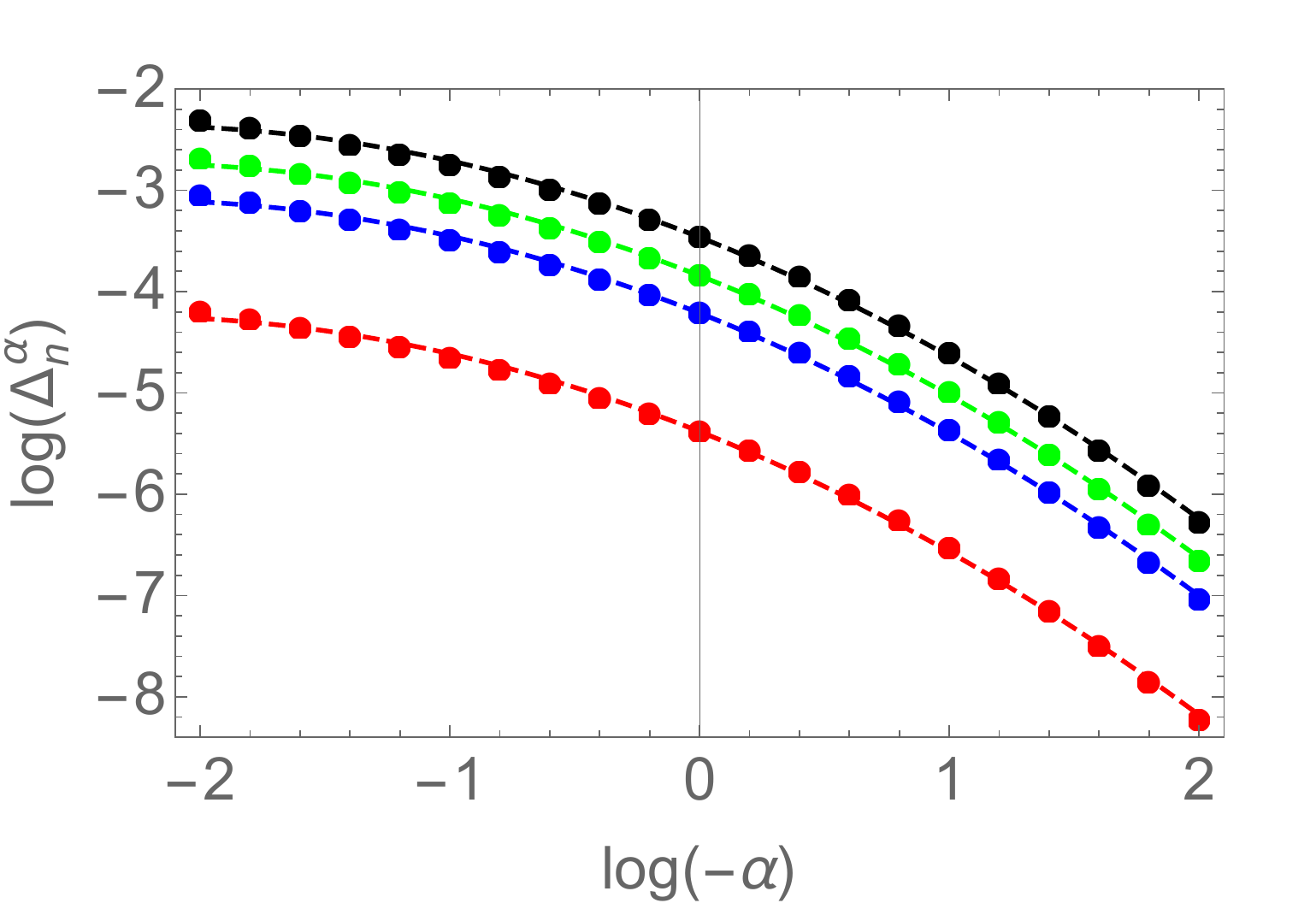}

				    \caption{The function (\ref{DIsing}) for $n=2$ (red), $n=5$ (blue), $n=7$ (green) and $n=10$ (black). The figure on the right shows clearly how all functions scale similarly with $\alpha$. We find that all functions are extremely well fitted by quadratic polynomials of the form $f(n)+k(n) x + p(n) x^2$ with $x=\log(-\alpha)$ which are presented as dashed lines.}
				     \label{figure2}
    \end{center}
    \end{figure}

In the absence of an analytic formula, the integral can be evaluated numerically for different values of $\alpha$ and $n$ as shown in Figs.~\ref{figure1} and \ref{figure2}. 
Fig. \ref{figure1} shows the function $\Delta_n^\alpha$ as a function of $n$ for different negative choices of $\alpha$. As observed for other fields in \cite{longpaper}, the values of $\Delta_n^\alpha$ decrease as the modulus of $\alpha$ increases. From the figure it is clear that for large $n$ the leading behaviour of  $\Delta_n^\alpha$ is linear in $n$, just as found in \cite{chin}. The curvature of the function is only visible for very small values of $n$, around 1. 
It is possible to expand (\ref{DIsing}) for large $n$ to work out that the leading contributions are:
\beq 
\Delta_n^\alpha = a(\alpha)+ n b(\alpha)+ \frac{1}{n} c(\alpha)+ \mathcal{O}(n^{-2})\;,
\eeq 
with 
\beq 
a(\alpha):=\frac{\alpha}{8\pi^2}\int_{-\infty}^\infty d\theta \frac{\theta^2 }{\pi^2+\theta^2} \frac{\sinh^2\frac{\theta}{2}}{\cosh\frac{\theta}{2}}\left|\frac{\sin(\frac{\alpha}{2}\sinh\theta)}{\frac{\alpha}{2}\sinh\theta}\right| \, e^{\frac{\alpha \theta}{2\pi}\sinh\theta}\;,
\eeq 
\beq 
b(\alpha):=\frac{1}{8\pi}\int_{-\infty}^\infty d\theta \frac{\theta }{\pi^2+\theta^2} \frac{\tanh\frac{\theta}{2}}{\cosh\frac{\theta}{2}}\left|\frac{\sin(\frac{\alpha}{2}\sinh\theta)}{\frac{\alpha}{2}\sinh\theta}\right| \, e^{\frac{\alpha \theta}{2\pi}\sinh\theta}\;,
\eeq 
and
\beq 
c(\alpha):=\frac{1}{192\pi^3 n}\int_{-\infty}^\infty d\theta \frac{\theta (\pi^4+\pi^2\theta^2-3\alpha^2\theta^2\sinh^2\theta) }{\pi^2+\theta^2} \frac{\sinh\frac{\theta}{2}}{\cosh^2\frac{\theta}{2}}\left|\frac{\sin(\frac{\alpha}{2}\sinh\theta)}{\frac{\alpha}{2}\sinh\theta}\right| \, e^{\frac{\alpha \theta}{2\pi}\sinh\theta}\;.
\eeq 
The coefficients of higher order corrections in $1/n$ can be easily evaluated by further expanding the integrand of (\ref{DIsing}).
As expected, $a(0)=0$ and $b(0)=-c(0)=\frac{1}{48}$.
If we assume that the function (\ref{DIsing}) still characterises the short-distance scaling of the two-point function $\bra \TT(0)\TT^\dagger(r)\ket_n^\alpha$ as well as the scaling of the vacuum expectation value $\bra \TT \ket_n^\alpha\sim m^{2\Delta_n^\alpha}$, then the results above already have important implications regarding the behaviour of the entropy. In order to understand this we need to briefly review the relationship between BPTF correlators and entanglement.
\section{Entanglement Entropy} 
The best-known measure of the entanglement between two parts of a quantum system, that is a bipartite measure, is the entanglement entropy or von Neumann entropy. Its good properties as an entanglement measure were established in \cite{bennet}. The simplest case in which it can be computed is that of a quantum model in a pure state $|\Psi\ket$. Let the Hilbert space of the theory be divided into two complementary subspaces $\mathcal{H}=\mathcal{H}_A\otimes \mathcal{H}_{\bar{A}}$ and let the reduced density matrix of subsystem $A$ be defined as
\beq 
\rho_A=\mathrm{Tr}_{\bar{A}}(|\Psi \ket \bra \Psi|)\;.
\eeq 
Then, the von Neumann entropy is defined as
\beq 
S(r)=-\mathrm{Tr}_A(\rho_A \log \rho_A)\;,
\eeq 
where $r$ is taken to be the length of subsystem $A$ in a 1D theory. Related to this is the family of R\'enyi entropies, defined as 
\beq 
S_n(r)=\frac{\log(\mathrm{Tr}_A(\rho_A^n))}{1-n}\;.
\eeq 
The normalisation of $\rho_A$ is chosen so that $\mathrm{Tr}_A(\rho_A)=1$, which implies that the von Neumann entropy can be obtained as a limit of the R\'enyi entropies
\beq
S(r)=\lim_{n\rightarrow 1} S_n(r)\;.
\eeq

In the context of critical systems it was realised a long time ago \cite{CallanW94,HolzheyLW94,Calabrese:2004eu} that for $n$ integer $\mathrm{Tr}_A(\rho_A^n)$ has a natural physical interpretation as a suitably normalised partition function in an $n$-sheeted Riemann surface with cyclically connected sheets. The structure of this Riemann surface is dictated by the product of $n$ matrices $\rho_A$ and by the trace, which induces the cyclicity property. The partition function picture applied to a configuration where the regions $A$ and $\bar{A}$ are both complementary and fully connected, is very advantageous in the context of conformal field theory. In this case a conformal map can be employed to map the Riemann surface back to the complex plane with conical singularities at the boundary point(s) between regions $A$ and $\bar{A}$. Henceforth both the R\'enyi and von Neumann entropies can be easily obtained leading to the well-known logarithmic growth of both $S(r)$ and $S_n(r)$ in critical systems \cite{CallanW94,HolzheyLW94, Latorre1, Jin, Calabrese:2004eu, Latorre2, Latorre3}.

Conformal maps can however not be used in massively perturbed and/or $\TTb$-perturbed CFTs and so an alternative approach to compute the partition function is needed. BPTFs provide such an approach. It can be shown that for a bipartition of the type described above, the function $\mathrm{Tr}_A(\rho_A^n)$ is proportional to a correlation function of BPTFs \cite{entropy,Calabrese:2004eu}. This correlation function contains as many fields as boundary points between $A$ and $\bar{A}$. In the simplest case, when both regions are semi-infinite, and $\bal=0$, we require only one field and we can write
\beq 
\mathrm{Tr}_A(\rho_A^n) = \varepsilon^{2\Delta^{\bol}_n} \bra \TT \ket_n^{\bol}\;,
\eeq 
where $\varepsilon$ is a non-universal short-distance cut-off and $\bra \TT \ket_n^{\bol}$ is the vacuum expectation value (VEV) of the BPTF. Plugging this expression into the formula for the R\'enyi entropies it is easy to derive the well-known property of saturation of the entanglement entropy in gapped systems \cite{Calabrese:2004eu,Hastings}, namely
\beq 
S(\infty)=-\frac{c}{6}\log(m\varepsilon)+\frac{U}{2}\;,
\label{gap}
\eeq 
where $U$ is related to the proportionality constant $v(n)$ in $\bra \TT \ket_n^{\bol}=v(n) m^{2\Delta^{\bol}_n}$ as
\beq 
U:=-\left.\frac{d v(n)}{dn}\right|_{n=1}\,\quad \mathrm{with}\quad v(1)=1\;.
\label{U}
\eeq 
On the other hand, if the region $A$ is finite of length $r$, the trace is given by
\beq 
\mathrm{Tr}_A(\rho_A^n) = \varepsilon^{4\Delta^{\bol}_n} \bra \TT(0)\TT^\dagger(r) \ket_n^{\bol}\;.
\eeq 
If the theory is critical, this formula can also be used to obtain the logarithmic growth of the entropies, which follows simply from power-law scaling of the two-point function. In CFT we have that 
\beq 
S_n(r)=\frac{c(n+1)}{6n} \log \frac{r}{\varepsilon},\quad S(r)=\frac{c}{3}\log \frac{r}{\varepsilon}\;,
\label{nongap}
\eeq 
where $c$ is the central charge. Crucially the pre-factor of the logarithmic terms in (\ref{nongap}) is related to the dimension of the BPTF in a simple way. For the R\'enyi entropies, this prefactor is given by $-\frac{4\Delta_n^{\bol}}{1-n}$ whereas for the von Neumann entropy it can be obtained as 
\beq 
-\lim_{n \rightarrow 1}\frac{4\Delta_n^{\bol}}{1-n}= 4\left.\frac{d \Delta_n^{\bol}}{dn}\right|_{n=1}\;.
\label{basics}
\eeq 
All formulae above have been written for the case $\bal=\bol$. Assuming that the relationship between correlators of BPTFs and entanglement measures is preserved by a generalised $\TTb$ perturbation,  it follows that if $\Delta_n^{\bal}$ has a different dependence on $n$ than $\Delta_n^0$, then the coefficient of the log-term in the entanglement entropy and R\'enyi entropies  will be modified. That is, both the function inside the limit and the limit itself in (\ref{basics}) will be different. Notice that for (\ref{basics}) to be correct, it is essential that $\Delta_1^{\bol}=0$, a condition that is a consequence of $\TT$ becoming the identity field in the absence of replicas. The $\Delta$-sum rule formulae presented in \cite{chin} violate this essential property. 

\subsection{Entanglement Entropy of a Finite Interval}
As we have just discussed
\begin{equation}
    S(r)= \lim_{n \to 1} S_n(r)=- \lim_{n \to 1} \; \frac{d}{dn} \Tr_{A}\rho_A^n = - \lim_{n \to 1} \; \frac{d}{dn} \varepsilon^{4 \Delta_n^\alpha} \langle \mathcal{T}(0) {\mathcal{T}^\dagger}(r) \rangle\;.
\end{equation}
The two-point correlation function of twist fields can be then calculated using its spectral representation in terms of form factors:
\begin{equation}
     \langle \mathcal{T}(0) {\mathcal{T}}^\dagger(r) \rangle_n^{\bal} = \sum_{k=0}^{\infty} \int_{-\infty}^{\infty} \frac{d \theta_1 \dots d \theta_k}{k! (2 \pi)^k} | F_k^{\mathcal{T}}(\theta_1, \dots, \theta_k;n,\bal) |^2 e^{-mr \sum_{i=1}^k \cosh\theta_i}\;.
\end{equation}
As is common with these kinds of expansions, we can truncate the sum at the two-particle level and so consider the leading non-trivial contribution for large $r$, which for the Ising model comes from the two-particle form factors. We can write
\beqa
     \langle \mathcal{T}(0) {\mathcal{T}}^\dagger(r) \rangle_n^{\bal} & =& (\langle \TT \rangle_n^{\bal})^2 + \sum _{i,j}^n \int_{-\infty}^{\infty} \int_{-\infty}^{\infty} \frac{d\theta_1 d \theta_2}{2! (2 \pi)^2} |F_2^{ij}(\theta_{12};n,\bal)|^2 e^{-mr(\cosh \theta_1 + \cosh \theta_2)} \nonumber\\
     &=& (\langle \TT \rangle_n^{\bal})^2 \left( 1 + \frac{n}{4 \pi^2} \int_{-\infty}^{\infty} d\theta g(\theta;n,\bal) K_0(2rm \cosh\frac{\theta}{2})\right)\;,
     \label{renyi}
\eeqa
where $K_0(z)$ is the modified Bessel function of the second kind and
\begin{flalign}
     (\langle \mathcal{T} \rangle_n^{\bal})^2  g(\theta;n, \bal) & = \sum_{j=1}^n |F_2^{1j}(\theta;n,\bal)|^2 =\\
     & =|F_2^{11}(\theta;n,\bal)|^2 + \sum_{j=1}^{n-1} |F_2^{11}(-\theta + 2 \pi i j;n, \bal)|^2\;.
     \label{sum}
\end{flalign}
The last identity above follows from the relations between form factors on different copies we presented earlier. Specialising again to the $\TTb$ deformation, we can use (\ref{twistFF}) to write
\beq 
|F_2^{11}(\theta;n,\alpha)|^2  =  \frac{(\langle \mathcal{T} \rangle_n^\alpha)^2 \sinh^2 \frac{\theta}{2n} \cos^2 \frac{\pi}{2n}}{n^2 \sinh^2{(\frac{i \pi -\theta}{2n})}\sinh^2{(\frac{i \pi+\theta}{2n})}} e^{\frac{\alpha m^2}{\pi} \frac{\theta}{n} \sinh{\theta}}\;,
\eeq
and
\beqa
&&|F_2^{11}(-\theta + 2 \pi i j;n,\alpha)|^2  =  F_2^{11}(-\theta + 2 \pi i j;n,\alpha) (F_2^{11})^*(-\theta - 2 \pi i j;n,\alpha) = \nonumber\\
&& = - \frac{i \langle \mathcal{T} \rangle_n^{\alpha} \sinh{(\frac{-\theta+2\pi i j}{2n})} \cos{(\frac{\pi}{2n})} }{n \sinh{[\frac{ i \pi +\theta - 2\pi i j }{2n}]} \sinh{[\frac{i \pi- \theta + 2\pi i j }{2n}]} } 
e^{-\frac{\alpha m^2}{2 \pi }(i \pi +\frac{\theta-2\pi i j}{n}) \sinh(-\theta+2\pi i j)} \times \nonumber\\
&& \times \frac{i \langle \mathcal{T} \rangle_n^{\bal} \sinh{(\frac{-\theta-2\pi i j}{2n})} \cos{(\frac{\pi}{2n})} }{n \sinh{[\frac{ -i \pi +\theta+2\pi i j }{2n}]} \sinh{[\frac{-i \pi- \theta-2\pi i j }{2n}]} } e^{-\frac{\alpha m^2}{2 \pi }(-i \pi +\frac{\theta+2\pi i j}{n}) \sinh(-\theta-2\pi i j)} =\nonumber\\
&&=
    \frac{ (\langle \mathcal{T} \rangle_n^\alpha)^2\sinh{(\frac{-\theta + 2 \pi ij}{2n})} \sinh{(\frac{\theta + 2 \pi ij}{2n})} \cos^2\frac{\pi}{2n} e^{\frac{\alpha m^2}{\pi n} \theta \sinh\theta} }{n^2 \sinh{(\frac{i \pi +\theta - 2 \pi ij}{2n})}\sinh{(\frac{i \pi-\theta + 2 \pi ij}{2n})} \sinh{(\frac{-i \pi +\theta + 2 \pi ij}{2n})}\sinh{(\frac{i \pi +\theta +  2 \pi ij}{2n})}} \;.
    \label{43}
\eeqa 
The integral in (\ref{renyi}) can be performed numerically for $n$ integer but due to the presence of sums such as in (\ref{sum}) it does not make sense to consider real, non-integer values of $n$. These are however needed in order to compute the derivative and limit required in the computation of the von Neumann entropy. In \cite{entropy} the analytic continuation to real positive $n\geq 1$ was performed using the cotangent trick, that is replacing the sum (\ref{sum}) by a contour integral in the variable $j$ with poles at $j=1,\ldots,n-1$. The same idea can be used here. The upshot of the computation in \cite{entropy} was that the sum has a special behaviour when $n\rightarrow 1$ and $\theta \rightarrow 0$ simultaneously which can be characterised by a $\delta$-function centered at $\theta=0$. The same behaviour occurs here since at $\theta=0$ the exponential factor, which is the only factor that depends on $\alpha$ is just 1. Both for the perturbed and unperturbed Ising model, it is possible to perform the sum 
\beq 
n \sum_{j=1}^{n-1} |F_2^{11}(2\pi i j; n, \alpha)|^2=\frac{n}{2}\quad \mathrm{for}\quad n>1\;. 
\eeq 
exactly. In fact, whether or not $\alpha=0$ the value of the sum is the same because the $\sinh\theta$ in the exponential factor is vanishing for $\theta=2\pi i j$. This result however reveals a non-analytic behaviour around $n=1$. In fact for $n=1$ the sum should be 0, but the formula above suggests that the analytic continuation from $n$ large to $n=1$ should give the value $1/2$ at $n=1$. It was shown in \cite{entropy} that the value $1/2$ at $n=1$ and the non-analyticity at $\theta=0$ are found for all diagonal IQFTs, even when they are interacting.

The similarity with the unperturbed case  extends fully to the computation of the sum (\ref{sum}) by contour integration. The same reasoning presented in Appendix 3 of \cite{entropy} applies here, crucially because the exponential factor in (\ref{43}) does not depend on the index $j$ which becomes the integration variable when using the cotangent trick. Therefore, we arrive to the interesting conclusion that the next-to-leading order correction to the entanglement entropy of a large interval is exactly the same as in the unperturbed theory {and} it does not depend on the perturbation, even if  the leading order term coming from the VEV (the saturation value) appears to be perturbation-dependent. We have then the same result as in \cite{entropy}, namely
\beq
S(r;\alpha)-S(\infty,\alpha)=-\frac{1}{8}K_0(2mr)+\mathcal{O}(e^{-3mr})\;,
\label{eq:SrminusSinfty}
\eeq 
where we now introduced the $\alpha$-dependence explicitly in the notation for $S(r;\alpha)$. 

This conclusion is not too surprising if we consider the physics of $\TTb$ perturbations. We know that for $\alpha>0$ ($<0$) we can interpret the effect of the perturbation as particles acquiring a positive (negative) length. This has strong effect on the short-distance (UV) physics of the theory. However, when distance is large compare to particle's width, this width should not play a major role and so the entropy correction remains unchanged. While this argument is qualitative in nature and not totally satisfactory, we believe it supports the correctness of the result \eqref{eq:SrminusSinfty}. We plan to investigate this point in more depth in the future.

The results of the $\Delta$-sum rule and of the general expansion of the two-point function as discussed in \cite{longpaper} -- and also applicable here -- both suggest that the leading contribution $S(\infty,\alpha)$ to the entanglement entropy is perturbation dependent, that is $\bra \TT\ket_n^\alpha \neq \bra \TT\ket_n^0$. This is also supported by the perturbative results obtained in \cite{Ashkenazi:2023fcn} and those of \cite{chin}. So even if the next-to-leading order correction is $\alpha$-independent, the leading contribution is not. 

{As} in the unperturbed case, the R\'enyi entropies acquire a less universal large-distance correction. In theories where the field $\TT$ has a non-vanishing one-particle form factor\footnote{In this case, our conclusion for the von Neumann entropy above will still hold, since the one-particle form factor is vanishing as $n\rightarrow 1$.}, the leading correction to the saturation of the R\'enyi entropies will be $\frac{n}{\pi} |F_1(n,\alpha)|^2$, which is very similar to the usual one (see e.g. \cite{mytoda,entropy4}), except that the one-particle form factor will be a function of $\alpha$. 
If the leading correction comes from the two-particle form factor, as in the Ising model, then as our formula (\ref{renyi}) shows, the integral will only be convergent for $\alpha<0$ due to the presence of an exponential factor $\exp({\frac{\alpha m^2}{n\pi} \theta\sinh\theta})$ and will clearly depend on $\alpha$.

We close this paper with a discussion of the leading term $S(\infty,\alpha)$ in \eqref{eq:SrminusSinfty}. This discussion is especially important because both in \cite{Ashkenazi:2023fcn} and in \cite{chin} it is claimed that $S(\infty,\alpha)$ is substantially different from the standard form (\ref{gap}) for the free fermion in the presence of a $\TTb$ perturbation. In both papers it is found that in addition to a logarithmic term -- just as in (\ref{gap}), albeit with a different coefficient -- there are terms of other types. In \cite{Ashkenazi:2023fcn} a term proportional to $\log^2{m \varepsilon}$ is found for the Dirac fermion and the same type of term is found in \cite{chin} for the Ising field theory. 

\begin{figure}[h!]
\begin{center}
	\includegraphics[width=7.5cm]{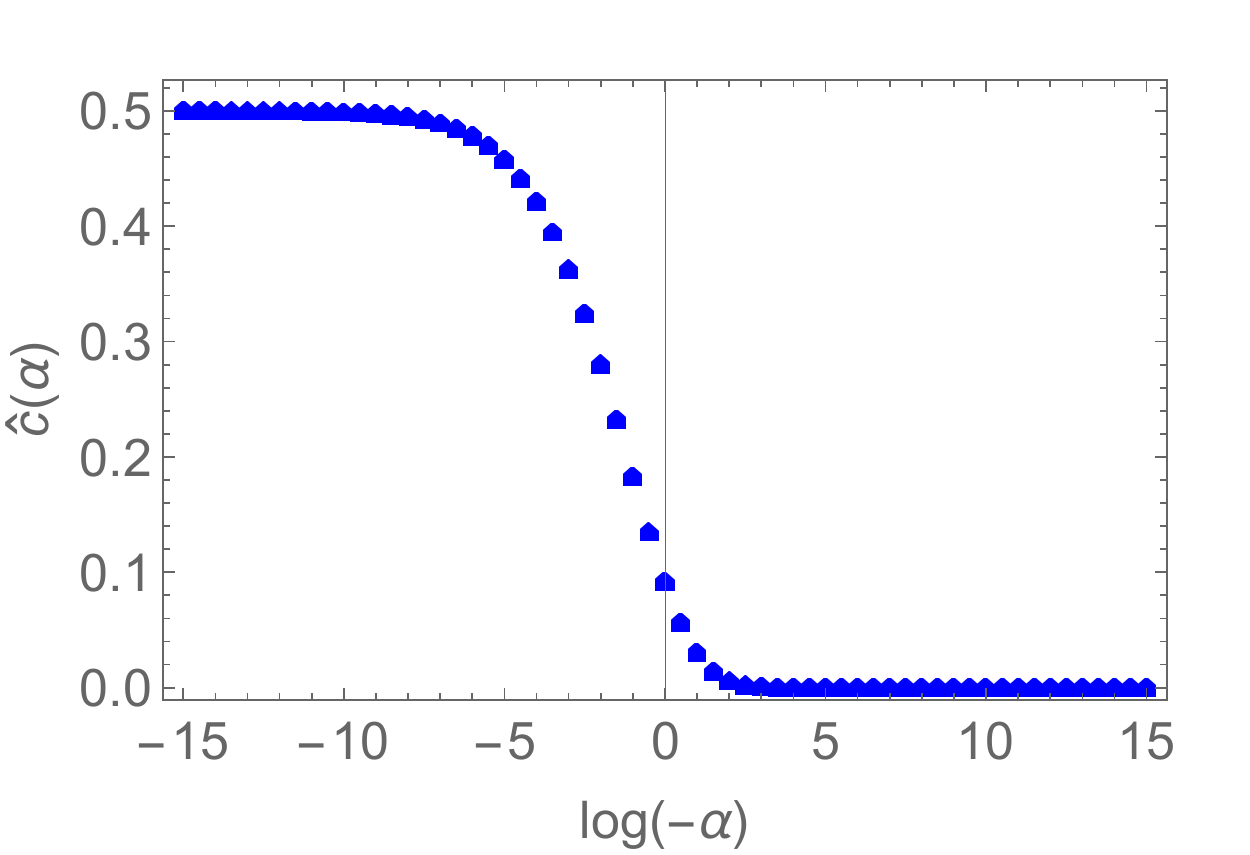}
 \includegraphics[width=7.5cm]{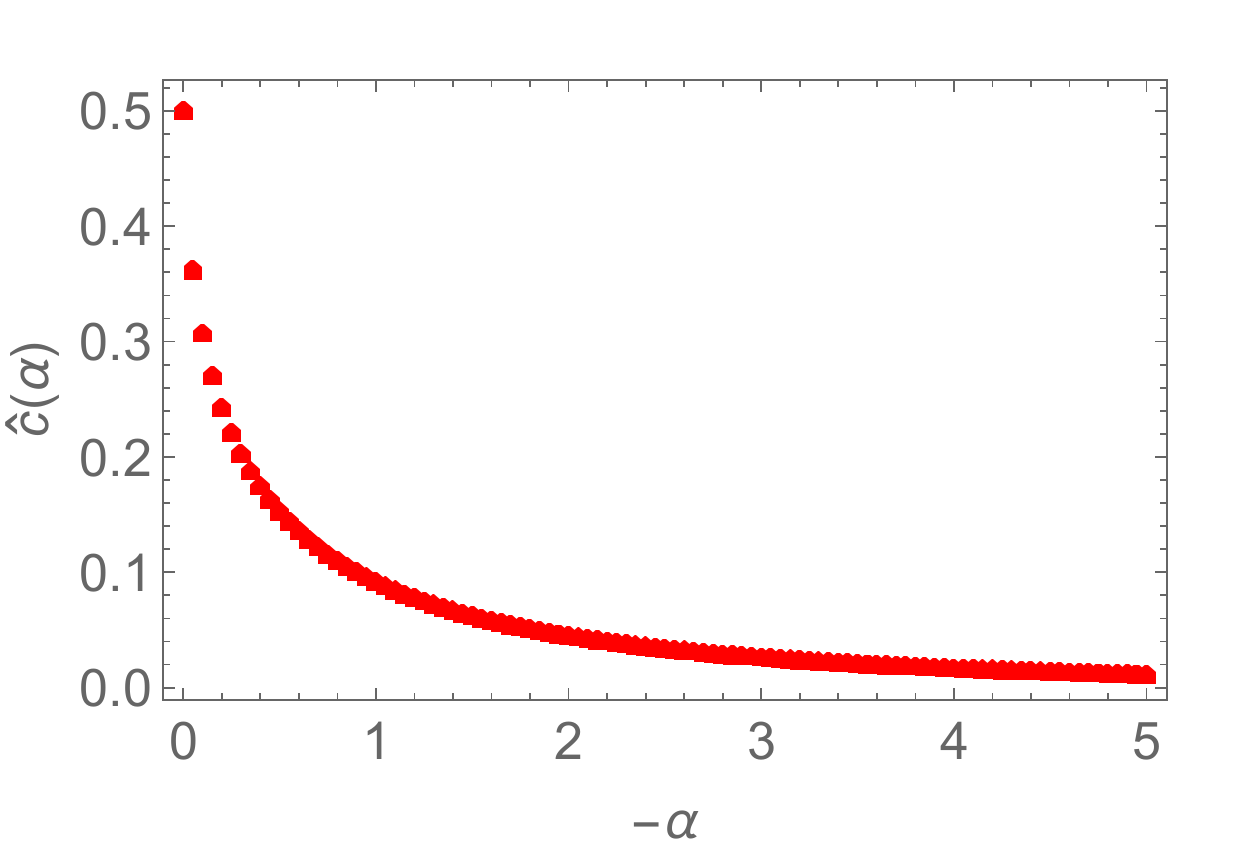}

				    \caption{The function $\hat{c}(\alpha)$ for $\alpha<0$ in the Ising field theory, as defined in (\ref{newc}). Although a distinct function from the function obtained from Zamolodchikov's $c$-theorem, the share the same main features. As expected $\hat{c}(0)=1/2$.}
				     \label{figure3}
    \end{center}
    \end{figure}

{Our calculation based on the $\Delta$-sum rule does not produce any of those additional, unusual terms, only a term proportional to $\log(m\epsilon)$ as usual, but with a different coefficient that we compute below. In fact, any computation based on the assumption that the vacuum expectation value of $\TT$ scales as a power of mass $m^{2\Delta_n^{\bal}}$ with $\Delta_{1}^{\bal}=0$ will inevitably produce a result of this type. The reason why \cite{chin} obtain something different is not entirely clear to us. It is evident though  that they assume a different type of scaling for the VEV. {In fact} the formula for $\Delta_n^{\bal}$ {used there} is not zero at $n=1$. As mentioned before, this is inconsistent with the integral form they use for $\Delta_n^{\bal}$, similar to our \eqref{DIsing}. Additionally, as we can see from (\ref{basics}) the equality between the limit and the derivative relies on the presence in $\Delta_n^{\bal}$ of a simple zero at $n=1$. If this were not true, as it is the case in \cite{chin},  the leading contribution to the von Neumann entropy would be  divergent due to the denominator $1-n$ of the R\'enyi entropies. In our case, using the formula (\ref{DIsing}), we have that the leading contribution to the entropy is given by
\beqa 
\lim_{n\rightarrow 1} \frac{\log \left(\varepsilon^{4\Delta_n^\alpha} (\bra \TT \ket_n^\alpha)^2\right)}{1-n}&=&\lim_{n\rightarrow 1} \frac{\log \left((m\varepsilon)^{4\Delta_n^\alpha} v(n,\alpha)^2\right)}{1-n}=\nonumber\\
&=&\log(m\epsilon) \lim_{n\rightarrow 1} \frac{4\Delta_n^{\alpha}}{1-n}+ \lim_{n\rightarrow 1} \frac{2\log v(n,\alpha)}{1-n}\;,
\label{calcu}
\eeqa 
where $v(n,\alpha)$ is the generalisation of the constant in (\ref{U}) and satisfies $v(1,\alpha)=1$. Since we do not know the explicit form of this function, we call 
\beq 
\lim_{n\rightarrow 1} \frac{2\log v(n,\alpha)}{1-n}=-\left.\frac{\partial}{\partial n} v(n,\alpha)\right|_{n=1}:=U(\alpha)\;.
\label{ual}
\eeq 
As for the main term we have that 
\beqa 
\lim_{n\rightarrow 1} \frac{4\Delta_n^{\alpha}}{1-n}=-4 \left.\frac{\partial \Delta_n^\alpha}{\partial n}\right|_{n=1}\;.
\eeqa
where from (\ref{DIsing})
\beq 
-4\left.\frac{\partial \Delta_n^\alpha}{\partial n}\right|_{n=1}=-\frac{1}{8}\int_{-\infty}^\infty d\theta \frac{\tanh^2\frac{\theta}{2}}{\cosh^2\frac{\theta}{2}}\left|\frac{\sin(\frac{\alpha}{2}\sinh\theta)}{\frac{\alpha}{2}\sinh\theta}\right| \, e^{\frac{\alpha \theta}{\pi}\sinh\theta}:=-\frac{\hat{c}(\alpha)}{3}\;.
\label{newc}
\eeq 
We can then write that, for the Ising field theory 
\beq 
S(r,\alpha)=-\frac{\hat{c}(\alpha)}{3} \log(m\varepsilon)+U(\alpha)-\frac{1}{8}K_0(2mr)+\mathcal{O}(e^{-3mr})\;,
\eeq 
with
\beq 
\hat{c}(0)=\frac{3}{8}\int_{-\infty}^\infty d\theta \frac{\tanh^2\frac{\theta}{2}}{\cosh^2\frac{\theta}{2}}=\frac{1}{2}\;.
\eeq 
As a function of $\alpha$,  $\hat{c}(\alpha)$ is different from the {one} obtained in \cite{PRL,longpaper} using Zamolodchikov's $c$-theorem \cite{Zamc}, even though both reduce to the correct value at $\alpha=0$ and display very similar dependence in $\alpha$ as shown in Figure~\ref{figure3} (this can be compared to figure 2 (right) in \cite{longpaper} and to figure 1 in \cite{PRL}).}
    
    Our results do not mean that unusual terms such as $\log^2(m\varepsilon)$ are not possible at all, so they do not necessarily contradict \cite{Ashkenazi:2023fcn}. Such terms could still be present. What our results show is that this would require the VEV of the BPTF to scale differently with the mass. VEVs are notoriously difficult to compute in QFT so this is a point that requires further investigation. A different scaling of the VEV is found in some simple unperturbed theories, like the non-compactified free boson (see \cite{entropy5,entropy6} for a detailed discussion), where the unusual scaling gives rise to a $\log\log(m\varepsilon)$ correction to the entropy. A similar type of correction appears naturally in computations of the symmetry resolved entanglement entropy \cite{GS,XAS} but in those cases the relevant function is not the VEV but its Fourier transform in a certain variable. 
More precisely, in order to obtain a correction proportional to $\log^2(m\varepsilon)$ it would be necessary to assume that
\beq 
\bra \TT \ket_n^\alpha=v(n,\alpha) m^{2\Delta_n^\alpha+\kappa(n,\alpha) (1-n)\log m}\;,
\eeq
with $\kappa(1,\alpha)\neq 0$, so that the calculation (\ref{calcu}) would give instead (leaving out the cut-off for now)
\beqa 
\lim_{n\rightarrow 1} \frac{\log (\bra \TT \ket_n^\alpha)^2}{1-n}
&=&\log m \lim_{n\rightarrow 1} \frac{4\Delta_n^{\alpha}+2\kappa(n,\alpha) (1-n)\log m}{1-n}+ \lim_{n\rightarrow 1} \frac{2\log v(n,\alpha)}{1-n}=\nonumber\\
&=& -\frac{\hat{c}(\alpha)}{3}\log m+ 2\kappa(1,\alpha) \log^2 m + U(\alpha)\;,
\label{calcu2}
\eeqa
with $U(\alpha)$ defined as in (\ref{ual}). We can then reintroduce the cut-off by replacing $m\mapsto m\varepsilon$ which is equivalent to adding a non-universal constant to (\ref{calcu2}) as well as making the coefficient of the $\log m$ term non-universal (it will become cut-off dependent as well). This is because 
\beqa 
&& -\frac{\hat{c}(\alpha)}{3}\log (m\varepsilon)+ 2\kappa(1,\alpha) \log^2 (m\varepsilon) + U(\alpha)\nonumber\\
 && = \left(-\frac{\hat{c}(\alpha)}{3}+2\log\varepsilon\right)\log m+ 2\kappa(1,\alpha) \log^2 m + U(\alpha) + d(\varepsilon)\,.
\eeqa 
with 
\beq 
d(\varepsilon)=2\kappa(1,\alpha)\log^2\varepsilon -\frac{\hat{c}(\alpha)}{3} \log\varepsilon\,.
\eeq 

\section{Conclusions and Outlook}
In this paper we have carried out a preliminary study of the form factors of branch point twist fields in generalised $\TTb$-perturbed theories. Our study is contemporary to and independent from the recent work \cite{chin}, even if the same questions are addressed. 

Our result for the minimal form factor of the branch point twist field agrees exactly with that of \cite{chin}. However, some of our conclusions regarding the $\Delta$-sum rule and the saturation value of the entropy in the Ising field theory are different from those reached in \cite{chin}. We believe this disagreement is due at least in part to an error in the evaluation of the $\Delta$-sum rule in \cite{chin} which we explained in more detail in the main section of this paper. 

Our main conclusion is the following. Starting with the two-particle form factor solution (\ref{22}) and assuming the $\Delta$-sum rule provides a reliable characterisation of the scaling of the vacuum expectation value of the branch point twist field, the von Neumann entropy of the Ising field theory will saturate to a value  which differs from that of the unperturbed theory. For a $\TTb$ perturbation, with $\alpha<0$, the saturation value is $-\frac{\hat{c}(\alpha)}{3}\log(\varepsilon m)+U(\alpha)$  with coefficients $\hat{c}(\alpha)$ and $U(\alpha)$ which are $\alpha$-dependent. The coefficient $\hat{c}(\alpha)$ reduces to $c$, the central charge of the UV unperturbed CFT for $\alpha=0$, but is smaller (in absolute value) for $|\alpha|>0$. Remarkably, the next-to-leading order correction to the von Neumann entropy remains unchanged with respect to the unperturbed theory, as shown also in \cite{chin}.

The results presented here and in our two previous papers \cite{PRL, longpaper}, open several new directions of investigation in the properties of irrelevantly deformed IQFTs, many of which were pointed out in \cite{longpaper}. The natural continuation of this work will involve the analysis of the higher particle form factors, a task on which we hope to be able to report in the near future.
\medskip

\noindent {\bf Acknowledgments:} The authors thank Benjamin Doyon and Michele Mazzoni for useful discussions. Olalla A. Castro-Alvaredo thanks EPSRC for financial support under Small Grant EP/W007045/1. The work of Stefano Negro is supported by the NSF grant PHY-2210349 and by the Simons Collaboration on Confinement and QCD Strings. Fabio Sailis is grateful for his PhD Studentship which is funded by City, University of London. 


\end{document}